\documentclass[twocolumn,times,tighten]{aastex631}

\newcommand{\Ho}{\mathrm{H}} 
\newcommand{\Ht}{\mathrm{H_2}}
\newcommand{\HHHp}{\mathrm{H_3^+}} 
\newcommand{\munan}[1]{{\color{black}#1}}
\newcommand{\kedron}[1]{{\color{black}#1}}

\newcommand{\kawai}[1]{{\color{black}#1}}

\usepackage{amsmath}
\usepackage{physics}
\usepackage{bbold}
\usepackage{soul}
\usepackage{booktabs} 
\usepackage{multirow} 
\usepackage{amsmath}  
\usepackage{lineno}
\usepackage{natbib}
\usepackage{comment}

\graphicspath{{./}{figures/}}

\begin{document}



\title{The cosmic ray ionization rate from H3+ observations can be overestimated due to neglect of time-dependent chemistry}
\correspondingauthor{Ka Wai Ho}
\email{kawaiho@kitp.ucsb.edu}

\author[0000-0003-3328-6300]{Ka Wai Ho}
\affiliation{Kavli Institute for Theoretical Physics, University of California, Santa Barbara, CA 93106, USA}
\affiliation{New Mexico Consortium, Los Alamos, NM 87544, USA} \affiliation{Department of Astronomy, University of Wisconsin--Madison, Madison, WI 53706, USA}

\author{Munan Gong} 
\affiliation{Department of Physics, University of Texas at El Paso, El Paso, TX 79968, USA} \affiliation{Max Planck Institute for Extraterrestrial Physics, Garching, Germany} 
\email{mgong2@utep.edu}

\author{Kedron Silsbee}
\affiliation{Department of Physics, University of Texas at El Paso, El Paso, TX 79968, USA}
\email{kpsilsbee@utep.edu}

\author{Alexei Ivlev} 
\affiliation{Max Planck Institute for Extraterrestrial Physics, Garching, Germany} \email{ivlev@mpe.mpg.de}

\begin{abstract}
The cosmic ray ionization rate (CRIR) is a key parameter governing the physical, chemical and thermal evolution of the interstellar medium. The primary technique for measuring the CRIR in diffuse molecular clouds relies on observations of ${\rm H_3^+}$. Previous analyses of these observations have derived the CRIR under the assumption of steady-state chemistry. Here, we investigate the effect of time-dependent chemistry on the inferred CRIR from ${\rm H_3^+}$ observations. We perform 3D MHD simulations with coupled chemistry and driven turbulence. Following procedures similar to those used in the literature to analyze ${\rm H_3^+}$ observations, we conduct mock CRIR measurements by post-processing our simulations with different values of the CRIR to obtain steady-state abundances of ${\rm H_2}$ and ${\rm H_3^+}$.  By comparing those with the abundances from time-dependent chemistry, we determine the best-fitting value of the CRIR. We find that the abundances of both ${\rm H_2}$ and ${\rm H_3^+}$ are higher in time-dependent chemistry simulations than in the steady-state case, especially in low-density regions. Furthermore, the inferred CRIR under the steady-state assumption is a factor of $\sim 2-5$ higher than the true CRIR, with a median value of $\zeta_\mathrm{inferred}/\zeta_\mathrm{true} \approx 3$. 
\munan{Using our fiducial model, we estimate an average CRIR per ${\rm H_2}$ of $\zeta_\Ht \approx 2\times 10^{-17}~\mathrm{s^{-1}}$ from the ${\rm H_3^+}$ observations.}
 \munan{The bias $\zeta_\mathrm{inferred}/\zeta_\mathrm{true}$} increases with stronger magnetic fields, weaker FUV radiation fields, and stronger turbulence. 
The CRIR is consistent with a constant value over the column density range of $N=(2-6)\times10^{21}~\mathrm{cm^{-2}}$.
\end{abstract}


\section{Introduction} \label{sec:intro}
Cosmic rays (CRs) are the primary source of ionization in dense molecular clouds and prestellar cores at values of $A_V$ greater than a few, where the interstellar ultraviolet radiation field is attenuated. CRs are involved in many of the important chemical reactions in the interstellar medium (ISM) by ionizing gas-phase molecules \citep{KIDA2010, Vasyunin2017}. On dust grains, CRs may catalyze reactions in the solid phase \citep{Shingledecker2018}, and facilitate transfer of material from the solid phase back to the gas phase \citep{HH1993}.  
In addition, by affecting the ionization fraction, the CRs regulate the rate of non-ideal MHD effects.  This affects the timescale for star formation \citep{Mouschovias2006}, the size and longevity of protoplanetary disks \citep{Zhao2016}, and the rate of dust coagulation \citep{Silsbee2020, Guillet2020}. Interestingly, many of the CR-induced effects are nearly proportional to the cosmic ray ionization rate (CRIR), and thus the influence of cosmic rays in the ISM is often parameterized by the CRIR of molecular hydrogen, $\zeta_{\rm H_2}$.

The first attempts to investigate the CRIR used near-Earth measurements of the energy spectrum of CRs, estimating a $\zeta_{\rm H_2} \approx 10^{-17} ~\mathrm{s^{-1}}$ \citep{Spitzer1968}.  In the subsequent decades, attempts to quantify the ionization rate have focused on chemical modeling of different molecular ions \citep{ODonnell1974, Black1977, Federman1996}. The CRIR has been inferred using $\rm H_3^+$, $\rm OH^+$, $\rm H2O^+$, $\rm ArH+$ in diffuse molecular clouds \citep{Indriolo2015, NW2017, Bacalla2019}, and using $\rm HCO^+$, $\rm DCO^+$, $\rm N_2H^+$, $\rm N2D^+$ in dense molecular cores \citep{Caselli1998, Redaelli2021, Pineda2024}. These measurements yield a value of CRIR in the range of $\sim 10^{-18} - 10^{-15}~\mathrm{s^{-1}}$. Recently, JWST observations of $\Ht$ rovibrational line emission provide a new method of measuring CRIR. However, this method is limited so far to one cloud, Barnard 68, with $\zeta_{\rm H_2} = 1.7\times 10^{-16} ~\mathrm{s^{-1}}$ \citep{Bialy2026}.

Among all the observational techniques of deriving CRIR from molecular ions, ${\rm H_3^+}$ has a particularly simple chemistry in diffuse molecular gas, and thus is a robust way to measure CRIR that is less subject to the uncertainties in chemical modeling \citep{McCall2003}.  \citet{Indriolo2007} used observations of $\rm H_3^+$ from absorption line spectroscopy to infer an average value of $\zeta_{\rm H_2} = 4.6 \times 10^{-16}$ s$^{-1}$ in a sample of 14 sightlines through diffuse molecular gas.  In doing so, they made a number of simplifying assumptions - namely that the gas density and temperature were constant along the line of sight, and that the free electrons in the gas are provided by carbon, which is assumed to be all in the form of C$^+$.  The gas density was estimated in most cases based on analysis of the relative level populations of C$_2$ \citep{Sonnentrucker2007}. \citet{IM2012} used similar methods as \citet{Indriolo2007}, and expanded the number of sightlines with $\rm H_3^+$ detections to 21, obtaining an average value of $\zeta_{\rm H_2} = 3.5 \times 10^{-16}$ s$^{-1}$.

To improve on the assumption of uniform chemistry and temperature along the line of sight in \citet{IM2012}, \citet{Neufeld2017} used a constant-density 1D PDR model with a slab geometry, accounting for variations in gas temperature and chemical abundances with the column density $N_\Ho$.  They then re-analyzed the 12 lines of sight which the gas density was estimated from C$_2$ level populations in \citet{IM2012}, and found an average value $\zeta_{\rm H_2} = 5.3 \times 10^{-16}$ s$^{-1}$, as well as evidence for very strong attenuation of the cosmic ray flux with increasing column density.

Recently, GAIA dust extinction measurements have enabled estimates of the 3D density structure in the local interstellar medium (ISM) \citep{Green2019, Lallement2022, Edenhofer2024}. Using the 3D gas density maps in \citet{Edenhofer2024}, \citet{Obolentseva2024} re-analyzed several sightlines using the 3D-PDR model \citep{Bisbas2012}, taking into account the spatial variations of FUV radiation field, temperature, and chemistry.  They found a reduction in the mean ionization rate by nearly a factor of 10, with an average of $\zeta_{\rm H_2} = 6 \times 10^{-17}$ s$^{-1}$. Most of this reduction arose because the density estimates from extinction mapping were considerably lower than previous estimates from C$_2$ level populations.  This conclusion was supported by the work on \citet{Neufeld2024}, who used updated calculations of the collisional de-excitation rates for rotational states of C$_2$ to revise many of the previous density estimates downwards. Expanding on \citet{Obolentseva2024}, \citet{Indriolo2026} increased the total number of sight lines from 7 to 16, obtaining a mean of $\zeta_{\rm H_2} = 5.3 \times 10^{-17}$ s$^{-1}$ with the standard deviation of $2.5 \times 10^{-17}$ s$^{-1}$.

One important aspect that has not been considered in the previous studies of CRIR measurements using $\rm H_3^+$ is the effect of time-dependent chemistry. In particular, $\Ht$, the precursor of $\rm H_3^+$, is likely to have non-steady-state abundances in diffuse atomic and molecular clouds where $\rm H_3^+$ is observed to constrain the CRIR. The formation timescale of $\Ht$ can be longer than the dynamical timescale in these environments \citep{GM2007, GOW2017, Gong2018, Gong2020_XCO, Joshi2019, Hu2021, Godard2023}. The current modeling of CRIR, however, assumes that $\Ht$ and $\rm H_3^+$ are both in chemical steady-state. We note that, even under the assumption of chemical equilibrium, the presence of strong compressive turbulence in the diffuse interstellar medium may affect H$_2$ abundance \citep{Bialy2019}. The origin of this effect, leading to systematically higher values compared to equilibrium, is the nonlinearity of H$_2$ self-shielding in local density perturbations produced by the turbulence. 

The goal of this paper is to investigate the effect of time-dependent chemistry on CRIR measurements from $\rm H_3^+$ by using magneto-hydrodynamic simulations of turbulent molecular clouds with chemistry. This paper is organized as follows. Section \ref{sec:Method} describes the method for our numerical simulations and CRIR mock observations. Section \ref{sec:Result} states the results from our analysis. Section \ref{sec:Discussion} discusses the implications of our results on the observations of CRIR. Finally, Section \ref{sec:conclusions} summarizes the main conclusions from this paper.

\section{Methods} \label{sec:Method}

\subsection{Numerical Simulations}
We conduct simulations with coupled MHD and chemistry using the publicly available Athena++ code \citep{Stone2020}. The chemistry module in Athena++ with self-consistent heating and cooling is implemented in \citet{Gong2023}. We perform 3D simulations of the multiphase ISM with driven turbulence using a Cartesian grid.
\subsubsection{Chemical Network}
We use the simplified carbon and oxygen network for atomic and molecular ISM in \citet{GOW2017}. This network is widely used and well tested, has been implemented in Athena++ \citep{Gong2023}, and is publicly available. It has the advantage of being relatively simple with a low computational cost, while still accurately capturing the main chemical and thermal processes in the atomic and molecular ISM. This network includes 18 atomic and molecular species of the elements $\mathrm{H}$, $\mathrm{He}$, $\mathrm{C}$, $\mathrm{O}$, and $\mathrm{Si}$, and 50 reactions. Heating and cooling processes from chemical reactions, FUV radiation, and atomic and molecular line cooling are included. The FUV radiation field was calculated using the six-ray radiation transfer module, which performs ray tracing along Cartesian grids to the edge of the simulation box.

The main creation pathway for $\mathrm{H_3^+}$, is cosmic ray ionization of $\mathrm{H_2}$ 
\begin{equation}\label{eq:rec_cr_H2}
    \mathrm{CR + H_2 \rightarrow H_2^+ + e},
\end{equation}
followed by reaction with a neutral H$_2$ molecule: 
\begin{equation}\label{eq:H2p_H2}
    \mathrm{H_2^+ + H_2 \rightarrow H_3^+ + e}.
\end{equation}
The ionization reaction \eqref{eq:rec_cr_H2} has rate coefficient $\zeta_\mathrm{H_2} = 2\zeta_p(2.3f_\mathrm{H_2} + 1.5f_\mathrm{H})$ \citep{GL1974}, where $\zeta_p$ is the primary CRIR per hydrogen nucleus. The reaction rate for \eqref{eq:H2p_H2} is fast, at $k_{\eqref{eq:H2p_H2}} = 2.84\times 10^{-9}T^{0.042}\exp(-T/46600)~\mathrm{cm^3 s^{-1}}$, where $T$ is temperature in Kelvin \citep{Linder1995, Glover2010}. Thus, almost every ${\rm H_2^+}$ that is created by \eqref{eq:rec_cr_H2} becomes an ${\rm H_3^+}$, unless the reaction occurs in predominately atomic gas with the molecular fraction less than $\sim30\%$ (${\rm H_2^+}$ can react with neutral $\Ho$ to form ${\rm H^+}$ at a slower rate of $6.4\times 10^{-10}~\mathrm{cm^3 s^{-1}}$ \citep{KIDA2010}).  
The destruction of $\HHHp$ is dominated by dissociative recombination with electrons that creates either $\mathrm{H_2 + H}$ or $\mathrm{3H}$
\begin{equation}\label{eq:rec_h3p}
    \mathrm{H_3^+ + e \rightarrow H_2 + H~or~3H}
\end{equation}
with the total rate coefficients for both branches $k_{\rm rec} = 1.3\times10^{-6}T^{-0.52}~\mathrm{cm^3 s^{-1}}$. Assuming $\HHHp$ is in steady-state (creation rate equal to destruction rate), and assuming that every ${\rm H_2^+}$ ion undergoes reaction \eqref{eq:H2p_H2}, the cosmic ray ionization rate of $\mathrm{H_2}$ can be written as
\begin{equation}\label{eq:k_CR}
    \zeta_\mathrm{H_2} = \frac{f_\HHHp}{f_\Ht}f_\mathrm{e}nk_{\rm rec},
\end{equation}
where $f_i$ is the number density of species $i$ normalized by the number density of hydrogen nuclei 
($f_i=n_i/n, n=n_\Ho + 2 n_{\rm H_2}$).
Equation \eqref{eq:k_CR} is often used to derive the CRIR in observations \citep{Indriolo2007, NW2017}. We note that in purely molecular gas ($f_\Ht=0.5)$,
\begin{equation}
    \zeta_\Ht=2.3\zeta_p.
\end{equation}

\subsubsection{Turbulence Driving}
To induce and maintain turbulence, we inject stochastic velocity perturbations in Fourier space using large-scale Ornstein–Uhlenbeck (OU) forcing, as commonly adopted in multiphase ISM simulations \citep{2018PhRvL.121b1104K, 2025PhRvD.112j1302H}. The forcing acts on modes with $k/k_{\rm min}\in[1.5,2.5]$ using a parabolic weighting that peaks at $k/k_{\rm min}=2$, where $k_{\rm min}\equiv 2\pi/L$ corresponds to the box scale. We construct the driving field as an equal mixture of solenoidal and compressive components with correlation time $t_{\rm corr}=2~{\rm Myr}$. The forcing amplitude is calibrated such that, in steady state, the velocity dispersion $u_{\rm rms} \approx 5.0~{\rm km,s^{-1}}$, in agreement with the observed velocity dispersion of the ISM on the length scale of $25\:{\rm pc}$ \citep{2023PASA...40...46K}.

\subsubsection{Initial and Boundary Conditions}
To represent diffuse molecular-cloud conditions, we initialize a cubic domain with a Cartesian grid of side length $L=50~{\rm pc}$. The initial gas density is uniform with number density of hydrogen nuclei $n=10~{\rm cm^{-3}}$, corresponding to a mean hydrogen column density $N=nL\simeq 1.5\times10^{21}~{\rm cm^{-2}}$. We impose an initial  uniform magnetic field $\mathbf{B}_0=B_0\,\hat{\mathbf{z}}$. The boundary condition is periodic in all directions for the MHD variables.  The six-ray approximation is used to model the external irradiation \citep{Gong2023}. We apply an incident FUV radiation field on all six faces of the domain and propagate the radiation inward along the Cartesian coordinates, attenuating the radiation field by dust- and molecular- shielding. We adopt a constant CRIR throughout the simulation domain.

The elemental abundances are set to be typical values of solar neighborhood ISM, with 0.1 for helium, $1.6\times 10^{-4}$ for carbon \citep{Sofia2004}, $3.2\times10^{-4}$ for oxygen \citep{SS1996}, and $1.7\times 10^{-6}$ for silicon \citep{Cardelli1994}. 
The initial temperatures and species abundances are set by evolving the chemistry and associated heating/cooling to equilibrium at the start of the calculation. We run each simulation for $26~{\rm Myr}$, corresponding to approximately $5$ times the eddies turnover time at the forcing scale, to ensure the turbulence is fully developed and in statistical steady state.

\subsubsection{Model Parameters}
\label{sec:param_study}

We run a set of simulations varying numerical and physical parameters, as listed in Table \ref{table:parameters}. All runs share the same box size, mean density, turbulence driving, boundary conditions, and chemical network described above.  We analyze the final snapshot of each run using the mock-observation and inference pipelines described in Sections~\ref{sec:mock_sightlines} and \ref{sec:mock_inference}.

Our fiducial model (FID) contains $256^3$ grid cells, corresponding to a spatial resolution of $\Delta x = L/256 = 0.195~{\rm pc}$. The fiducial value of CRIR and FUV radiation field is chosen to be $\zeta_p=2\times10^{-17}~\mathrm{s^{-1}}$ and $\chi=0.5$, corresponding roughly to the typical values derived by \citet{Obolentseva2024, Indriolo2026}. The fiducial initial magnetic field strength is $B_0=1\,\mu\mathrm{G}$, and the magnetic field strength at the end of the simulation reaches 
$\sqrt{\langle B^2 \rangle} \approx 4.66\,\mu\mathrm{G}$, which lies within the observed few-$\mu\mathrm{G}$ range of Galactic ISM magnetic fields \citep{HeilesTroland2005,Crutcher2010,BeckWielebinski2013}.
We additionally perform a lower-resolution run (LRES) with $128^3$ grid cells ($\Delta x=0.391~{\rm pc}$) to assess the numerical convergence. Note that even for LRES, the numerical resolution is still higher than the resolution of the dust map used to derive the CRIR in \citep{Obolentseva2024, Indriolo2026}.
The HB, HUV, HCR and HT runs probe environmental variations: HB increases the initial magnetic field from $B_0=1\mu G$ to $5~\mu{\rm G}$; \texttt{HUV} increases the incident FUV radiation field strength from $\chi=0.5$ to $\chi=3$, where $\chi$ is the UV field in units of the Draine field \citep{Draine1978}; the \texttt{HCR} run increases the CRIR by an order of magnitude; and finally HT increases the injection energy of turbulence, so that the steady state turbulent velocity dispersion reaches $8.21~\mathrm{km/s}$.

\begin{table}[htbp]
    \centering
    \small 
    \setlength{\tabcolsep}{4pt} 
    \caption{Model parameters for the simulation suite.}
    \label{table:parameters}
    \begin{tabular}{lccccc}
        \tableline
        \tableline
        Model ID & Resolution & $\zeta_p~[{\rm s^{-1}}]$ & $\chi$ & $B_0~[\mu{\rm G}]$ & $\sigma_v~[{\rm km~s^{-1}}]$\\
        \tableline
        FID  & $256^3$ & $2\times10^{-17}$ & 0.5 & 1 & 5.01 \\
        LRES & $128^3$ & $2\times10^{-17}$ & 0.5 & 1 & 4.80 \\
        HB   & $256^3$ & $2\times10^{-17}$ & 0.5 & 5 & 4.94 \\
        HUV  & $256^3$ & $2\times10^{-17}$ & 3   & 1 & 4.70 \\
        HCR  & $256^3$ & $2\times10^{-16}$ & 0.5 & 1 & 4.99 \\
        HT   & $256^3$ & $2\times10^{-17}$ & 0.5 & 1 & 8.21 \\
        \tableline
        \tableline
    \end{tabular}
\end{table}

\subsection{CRIR Mock Measurements}

\label{sec:mockcrir}
Recent ${\rm H_3^+}$-based determinations of $\zeta_{\rm H_2}$ rely on forward modeling of the full line-of-sight structure rather than the single-zone steady-state estimator in Eq.~\eqref{eq:k_CR}.
In particular, \citet{Obolentseva2024, Indriolo2026} combine parsec-resolution 3D dust maps constraining local density structure along each sightline with 3D-PDR calculations to predict ${\rm H_2}$ and ${\rm H_3^+}$ column densities for a grid of ionization rates. They infer $\zeta_{\rm p}$ by identifying the value that best matches the observed ratio of $N_{\rm H_2}$ and $N_{\rm H_3^+}$.
To investigate the systematic error in the observations from assumptions of steady-state chemistry, we follow the same methodology as in \citet{Obolentseva2024, Indriolo2026}, and perform ``mock observations'' of our simulated cloud by (i) post-processing our simulations with steady-state chemistry and a range of CRIR to construct ${\rm H_2}$ and ${\rm H_3^+}$ column densities along many pencil-beam sight-lines throughout the simulation volume and (ii) obtaining an ``inferred'' $\zeta_{\rm H_2}$ for each sight-line by choosing the steady-state $N_{\rm H_3^+}/N_{\rm H_2}$ that best matches the ``real'' time-dependent value in our original simulations. This inferred CRIR can then be directly compared to the true (input) CRIR used in the time-dependent simulations.

\begin{figure*}[t!]
    \centering
    \includegraphics[width=1.0\textwidth]{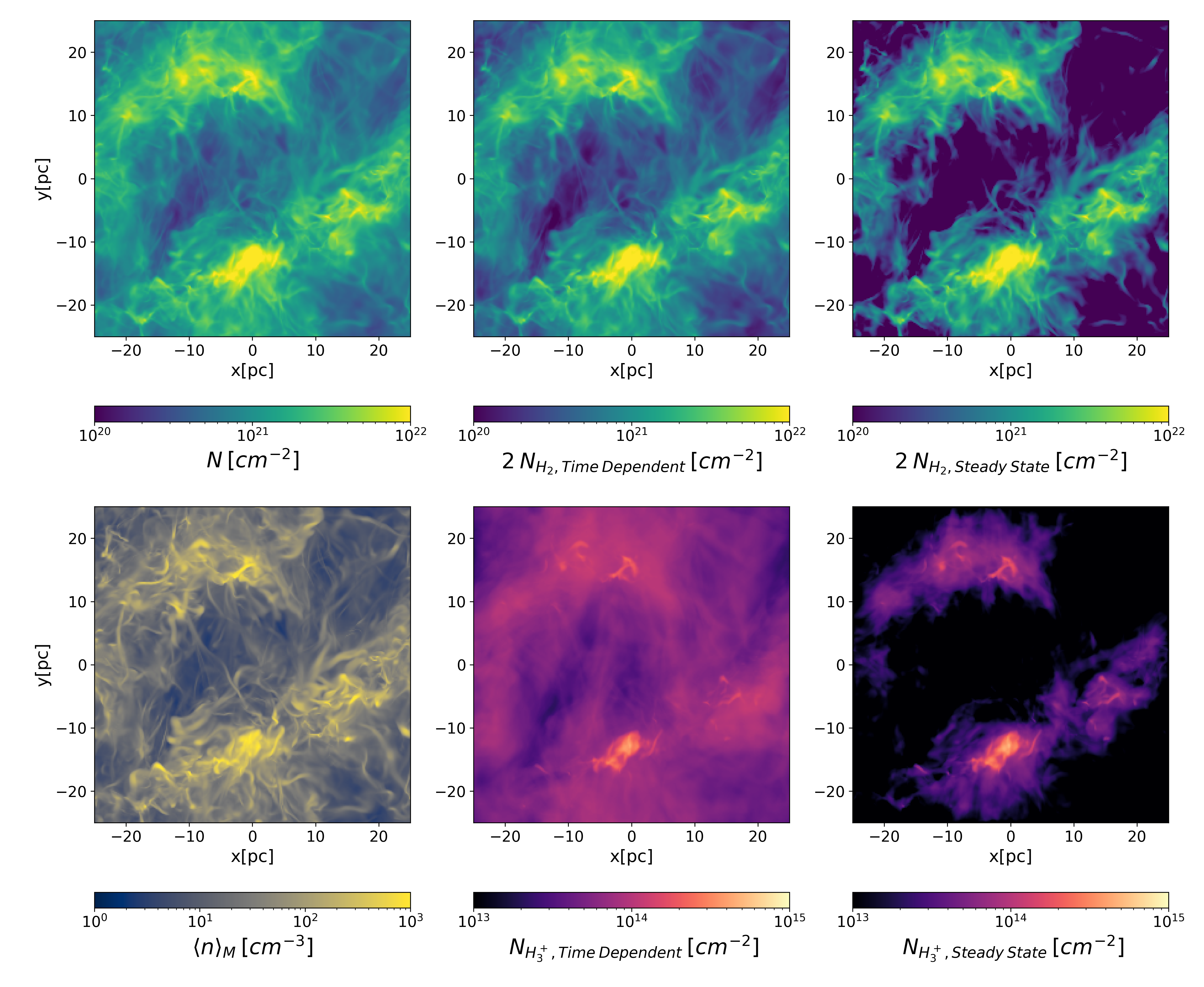}
    \caption{Column density maps comparing simulations with time-dependent and steady-state chemistry with the same CRIR.
    Top panels: total column density $N_{\rm H}$ (left), molecular column $2N_{\rm H_2}$ in the time-dependent chemistry simulation (middle), and the steady-state $2N_{\rm H_2}$ from post-processing (right).
    Bottom panels: mass-weighted mean density along the line of sight $\langle n\rangle_M$ (left), $N({\rm H_3^+})$ in the time-dependent chemistry simulation (middle), and the steady-state $N({\rm H_3^+})$ from post-processing (right). Both $\Ht$ and $\rm H_3^+$ exist at lower density/column density regions in time-dependent chemistry simulations compared to their steady-state values.}
    \label{fig:map}
\end{figure*}

\subsubsection{Post-processing with Steady-state Chemistry}\label{subsubsec:post-processing}
Following the method of deriving steady-state chemical abundances from density structures in the analysis of $\rm H_3^+$ observations \citep{Obolentseva2024, Indriolo2026}, we construct steady-state chemical abundances by post-processing our simulations with time-dependent chemistry. 
We select the final simulation snapshot and hold the gas density and velocity fixed. We then integrate the chemical rate equations, temperature evolution, and radiation transfer forward in time for $1~{\rm Gyr}$ to ensure that the chemistry has reached a steady-state with no advection of chemical species or dynamical evolution of the gas. We adopt the same incident FUV field strength as in the time-dependent simulation, and vary the CRIR of molecular hydrogen over the range $\zeta_{p}\in[2\times10^{-18},\,5\times10^{-16}]~{\rm s^{-1}}$, sampled at 12 evenly logarithmically spaced values. For the HCR run, we additionally extend the post-processing range of $\zeta_p$ up to $2.43\times10^{-15}~{\rm s^{-1}}$.
For each choice of $\zeta_{p}$, we obtain the steady-state  ${\rm H_2}$ and ${\rm H_3^+}$ abundances used in our mock CRIR inference (Section~\ref{sec:mock_inference}).

\subsubsection{Synthetic Sight Lines and Column Densities}
\label{sec:mock_sightlines}
We treat each pencil-beam sight-line through the computational domain as an independent synthetic observation.  We calculate the column density of ${\rm H_2}$ and ${\rm H_3^+}$ along directions parallel to each coordinate axis.  For example, parallel to the x-axis, the column density of species $s$ is
\begin{equation}
    N_s^{x}(y, z) \;=\; \int_{x' = 0}^{L} n_s(x', y, z)\,dx' 
\end{equation}

For a grid with $N_{\rm cell}^3$ resolution elements, this results in $3 N_{\rm cell}^2$ separate column density measurements for each chemical species.  

For each sight line we record $N^{\rm TD}_{\rm H_2}$ and $N^{\rm TD}_{\rm H_3^+}$ from the \emph{time-dependent} simulation output. We then compute the same set of column densities for each of the post-processed outputs (one for each trial value of $\zeta_{\rm H_2}$), obtaining the \emph{steady-state} abundances $N^{\rm SS}_X(\zeta_{\rm H_2})$.

\begin{figure*}[t!]
    \centering
    \includegraphics[width=1.0\textwidth]{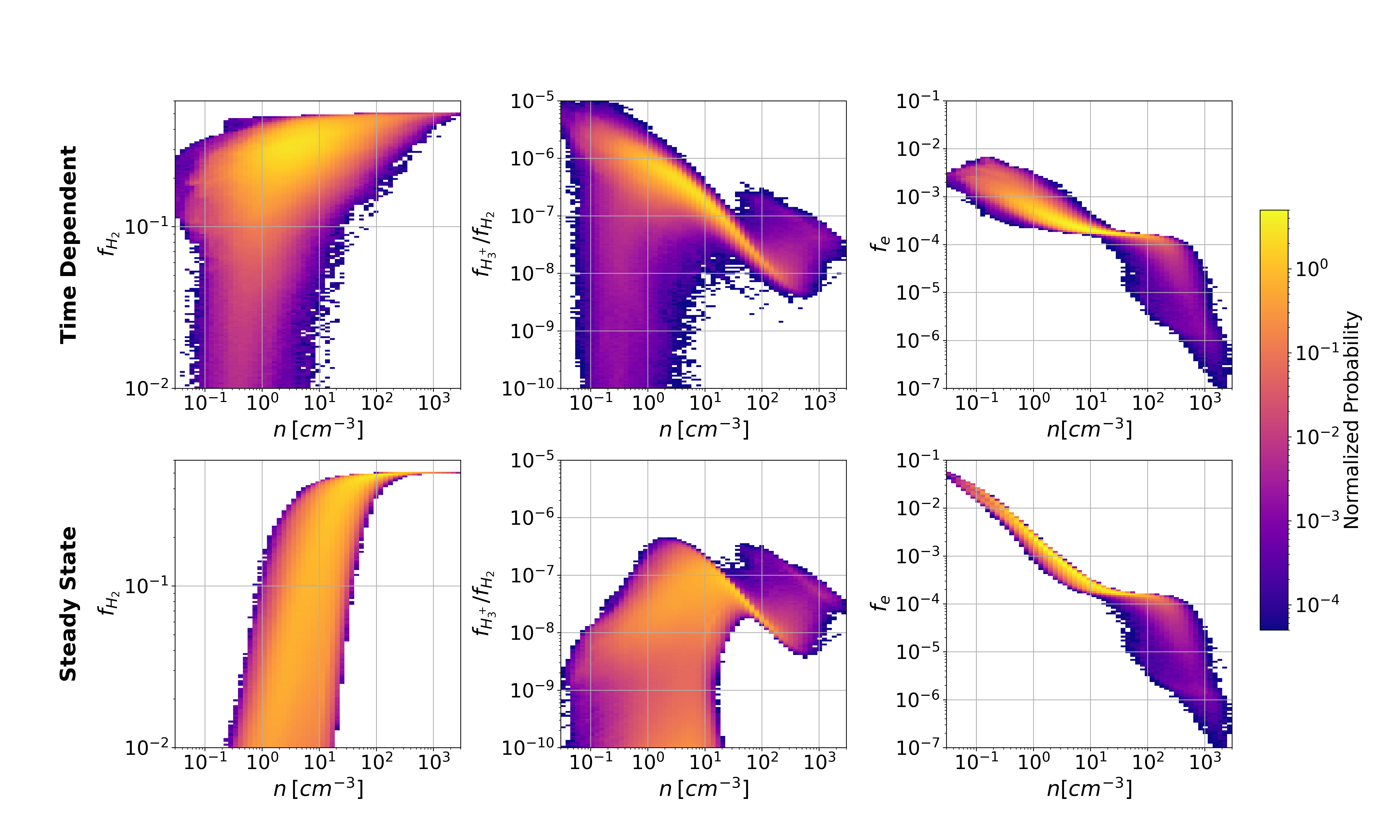}
    \caption{
    2D histograms of chemical states in the fiducial simulation snapshot shown in Figure \ref{fig:map} for time-dependent (top panels) and steady-state chemistry (bottom panels).
    Left: $\Ht$ fractional abundance $f_{\rm H_2}$ versus density $n$.
    Middle: abundance ratio $f_{{\rm H_3^+}}/f_{{\rm H_2}}$ versus density $n$.
    Right: electron abundance $f_e$ versus density $n$. With time-dependent chemistry, both the $\Ht$ abundance $f_{\rm H_2}$ and the abundance ratio $f_{{\rm H_3^+}}/f_{{\rm H_2}}$ are increased at lower densities $(n\lesssim 100~\mathrm{cm^{-3}})$ compared to steady-state chemistry. 
    }
    \label{fig:2dhist}
\end{figure*}

\subsubsection{Mock Inference of $\zeta_{\rm H_2}$}
\label{sec:mock_inference}
For each sight line, we obtain an ``inferred'' CRIR by comparing the time-dependent abundances to the steady-state values.
Specifically, we define a mismatch function in log-space,
%
\begin{equation}
\mathcal{D}(\zeta_{\rm p}) \;=\;
\left[\log \left( \frac{N^{\rm eq}_{\rm H_3^+}(\zeta_{\rm p})}{ N^{\rm eq}_{\rm H_2}(\zeta_{\rm p})} \right)-\log \left( \frac{N^{\rm TD}_{\rm H_3^+}}{ N^{\rm TD}_{\rm H_2}} \right)\right]^2
\label{eq:mismatch}
\end{equation}

and take the inferred CRIR $\zeta_{\rm inferred}$ to be the value that sets
\begin{equation}\label{eq:zeta_inferred}
\mathcal{D}(\zeta_{\rm inferred}) = 0.
\end{equation}

This inferred $\zeta_\mathrm{inferred}$ assuming steady-state chemistry can then be compared to the true CRIR $\zeta_\mathrm{true}$ used in our time-dependent chemistry simulations to assess the bias in modeling the observed sightlines in \citet{Obolentseva2024} and  \citet{Indriolo2026}. As done in \citet{Obolentseva2024} and \citet{Indriolo2026}, for each sight-line, we use bicubic interpolation of the data derived from our post-processing simulations to obtain $N_\HHHp/N_\Ht$ as a function of $\zeta_p$. We restrict our analysis to sight-lines with the total hydrogen column $1.6\times 10^{21}\,{\rm cm^{-2}} \leq N \leq 6.7\times 10^{21} \,{\rm cm^{-2}}$ the range of column densities where both $\Ht$ and $\HHHp$ are detected in observations\citep{Obolentseva2024, Indriolo2026}.

\section{Results} \label{sec:Result}
\subsection{Time-dependent versus Steady-state Chemistry}
\label{sec:td_vs_eq}
The comparison between time-dependent and steady-state chemical abundances for $\Ht$ and $\rm H_3^+$ is shown in Figure~\ref{fig:map}. The same CRIR is used in the post-processed steady-state solution as in the time-dependent chemistry simulation. The main difference appears in the diffuse gas: with time-dependent chemistry, the column densities $N({\rm H_2})$ and $N({\rm H_3^+})$ are significantly higher at gas columns $N \sim 10^{20}-10^{21}~\mathrm{cm^{-2}}$ and mass-weighted densities $\langle n \rangle_M \sim 1-100~\mathrm{cm^{-3}}$ compared to the steady-state chemical abundances. This can be understood by comparing the $\Ht$ formation timescale $t_\Ht \approx 10 \times  (n/100~\mathrm{cm^{-3}})^{-1}~\mathrm{Myr}$ to the dynamical timescale for turbulence mixing $t_\mathrm{dyn}\approx (L/\mathrm{pc})^{1/2}~\mathrm{Myr}$, resulting in a criterion of 
\begin{equation}
    \left( \frac{n}{100~\mathrm{cm^{-3}}} \right ) \left( \frac{N}{10^{21}~\mathrm{cm^{-2}}} \right ) > 31
\end{equation}
for $t_\Ht < t_\mathrm{dyn}$ (Equation (18) in \citet{GOW2017}). Therefore, we expect the $\Ht$ abundance to be out of steady-state in diffuse molecular gas. The $\rm H_3^+$ formation time, on the other hand, $t_{\rm H_3^+} \sim f_\HHHp/k_\mathrm{CR}\sim 10^{-7}/(5\times 10^{-17}~\mathrm{s^{-1}}) \sim 100~\mathrm{yr}$ is very short. Thus, $\HHHp$ reaches steady-state quickly with $\Ht$, and the distribution of $\HHHp$ closely follows that of $\Ht$ in Figure \ref{fig:map} for both time-dependent and steady-state chemistry.

Why is the $\Ht$ abundance higher with time-dependent chemistry in diffuse molecular gas than with steady-state chemistry? This arises from the nonlinearity of $\Ht$ self-shielding, which decreases the photo-dissociation rate steeply as the column density of $\Ht$ increases above $N(\Ht)\gtrsim 10^{19}~\mathrm{cm^{-2}}$ \citep{DB1996}. When $\Ht$ gas is brought from dense to diffuse gas by turbulent advection, it can be protected from photo-dissociation by self-shielding, enabling $\Ht$ to survive longer than the dynamical timescale in low-density regions in which it would not form efficiently with steady-state chemistry. The same phenomenon of increased $\Ht$ abundance in diffuse molecular clouds with time-dependent chemistry has been seen in other simulations \citep{Joshi2019, Hu2021, Godard2023}.

In dense gas with $n\gtrsim100~\mathrm{cm^{-3}}$ and $N\gtrsim 10^{21}~\mathrm{cm^{-2}}$, the chemical timescale is comparable or shorter than the dynamical timescale, and the time-dependent chemical abundances approach the steady-state solution.



\begin{figure}[tbp]
    \centering
    \includegraphics[width=0.5\textwidth]{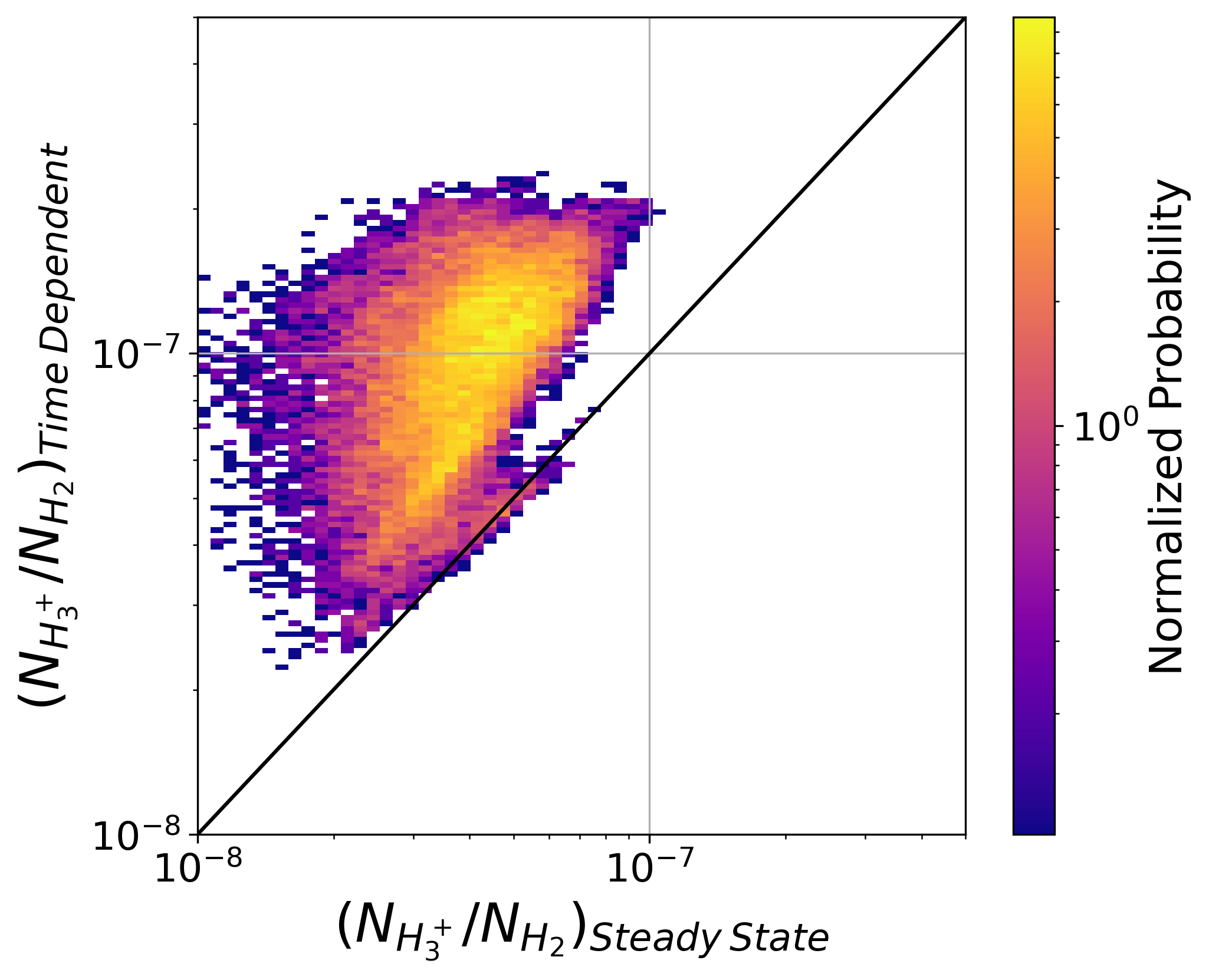}
    \caption{
    Comparison of the sight-line integrated column ratios $N_{\rm H_3^+}/N_{\rm H_2}$ between time-dependent and steady-state chemistry solutions for the fiducial simulation snapshot shown in Figures \ref{fig:map} and \ref{fig:2dhist}. The black diagonal line shows the one-to-one relation.}
    \label{fig:NH3_over_NH2}
\end{figure}

To further quantify the differences between time-dependent and steady-state chemistry, we show the 2D histograms of chemical states in Figure \ref{fig:2dhist}. At lower densities ($n\lesssim 100~\mathrm{cm^{-3}}$), not only the $\Ht$ abundance $f_{\rm H_2}$, but also the $\HHHp$ abundance relative to $\Ht$, $f_{\rm H_3^+}/f_{\rm H_2}$, is significantly higher with time-dependent chemistry. This is because $\HHHp$ forms through reactions between CR-produced $\rm H_2^+$ and $\Ht$, which are more efficient in the time-dependent case where $\Ht$ remains abundant at lower densities. In more atomic gas with $f_\Ht \lesssim 0.1$, $\rm H_2^+$ instead preferentially reacts with atomic $\Ho$ to form $\rm H^+$ rather than reacting with $\Ht$ to form $\HHHp$ in Equation \eqref{eq:H2p_H2} \citep{GOW2017}.
Another factor contributing to the higher $f_{\rm H_3^+}/f_{\rm H_2}$ in time-dependent chemistry is the lower effective recombination rate $nf_ek_\mathrm{rec}$ (Equation \eqref{eq:rec_h3p}), caused by both the lower densities at which $\Ht$ survives and the lower electron abundance there (left and right panels of Figure \ref{fig:2dhist}). We note that the gas temperature is somewhat lower with time-dependent chemistry in low-density regions due to increased $\rm H_2$ cooling, which increases $k_\mathrm{rec}$.  However, this does not offset the effect of decreased density and electron abundance enough, and the effective recombination rate $nf_ek_\mathrm{rec}$ is still lower. This directly leads to higher $N_\HHHp/N_\Ht$ ratios when integrated along each sight-line (Figure \ref{fig:NH3_over_NH2}). Consequently, the CRIR derived from steady-state chemistry assumptions is higher than the true value, as detailed in the next section.

\subsection{Impact of Time-dependent Chemistry on the inferred CRIR}
\label{sec:crir_bias}

\begin{figure}[tbp]
    \centering
    \includegraphics[width=0.5\textwidth]{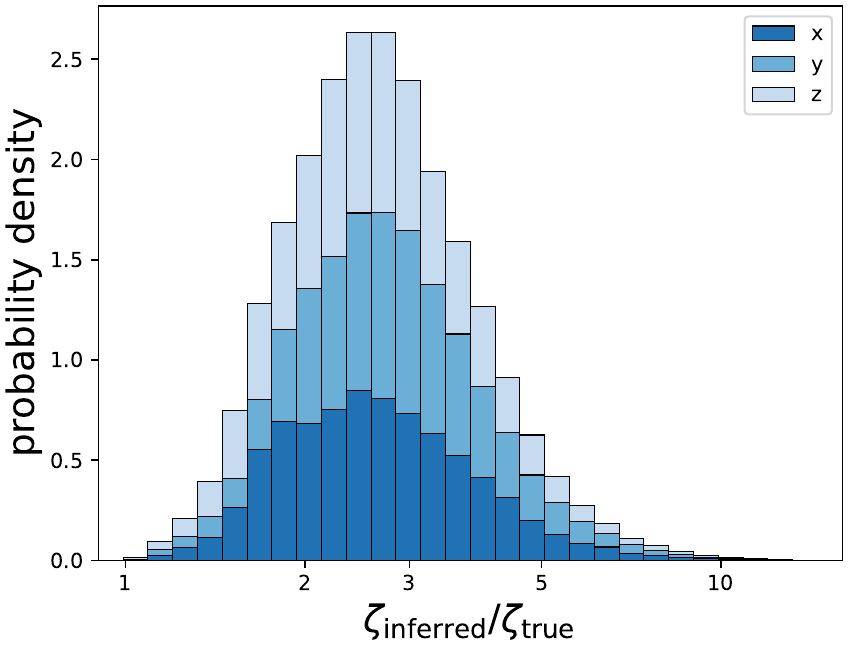}
    \caption{
    Histogram of the ratio of inferred to true CRIR $\zeta_{\rm inferred}/\zeta_{\rm true}$ for the fiducial model.  The different colors indicate sightlines parallel to different axes and shown in the legend. The inferred CRIR, assuming steady-state chemistry, overestimates the true CRIR by a factor of $\sim 2-5$.}
    \label{fig:zeta_hist}
\end{figure}

We next quantify how time-dependent chemistry affects ${\rm H_3^+}$-based CRIR measurements when the data are interpreted under the steady-state assumption.
Figure \ref{fig:zeta_hist} shows a histogram of the ratio of $\zeta_\mathrm{inferred}$ derived from mock inferences assuming steady-state chemistry (Equation \eqref{eq:zeta_inferred}) to the true CRIR $\zeta_\mathrm{true}$ used in our time-dependent chemistry simulations for all the sight-lines with $1.6\times 10^{21}\,{\rm cm^{-2}} \leq N \leq 6.7\times 10^{21} \,{\rm cm^{-2}}$. Overall, $\zeta_\mathrm{inferred}$ is larger than $\zeta_\mathrm{true}$ by a factor of $\sim 2-5$, with a median of $\zeta_\mathrm{inferred}/\zeta_\mathrm{true}\approx 3$. 
We do not see a significant difference between the sight-lines in different directions. The initial weak B-field along the z-axis does not introduce a significant amount of anisotropy, since the gas is overall super-Alfvenic with plasma $\beta$ $\approx 19$.


\begin{figure}[tbp]
    \centering
    \includegraphics[width=0.5\textwidth]{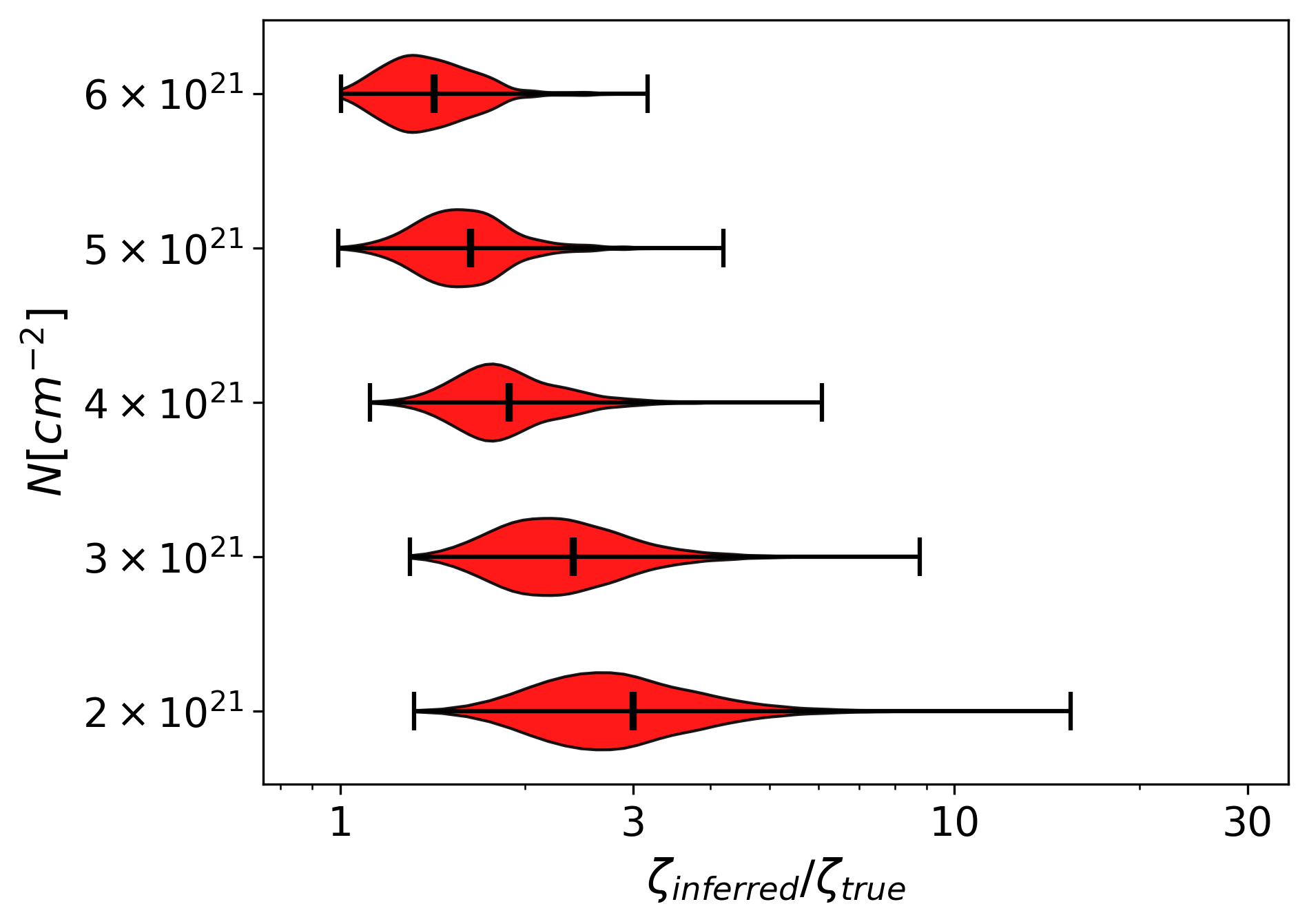}
    \caption{
    Violin plot illustrating the distribution of $\zeta_{\rm inferred}/\zeta_{\rm true}$ for different bins of column density $N$.  The data come from the fiducial model.. The black vertical segment marks the median of each distribution. The over-estimation of the CRIR, $\zeta_{\rm inferred}/\zeta_{\rm true}$, is higher at lower $N$.}
    \label{fig:violin}
\end{figure}

Furthermore, the bias in observational analysis assuming steady-state chemistry, $\zeta_{\rm inferred}/\zeta_{\rm true}$, has a dependence on the line-of-sight column density as shown in Figure~\ref{fig:violin}. As the column density decreases, the bias $\zeta_{\rm inferred}/\zeta_{\rm true}$ is larger, because the time-dependent chemical abundances of $\Ht$ and $\HHHp$ are further departed from the steady-state value (Section \ref{sec:td_vs_eq}). 

\subsection{Dependence on physical conditions}
\label{sec:param_dependence}
\begin{figure*}[tbp]
    \centering
    \includegraphics[width=1.0\textwidth]{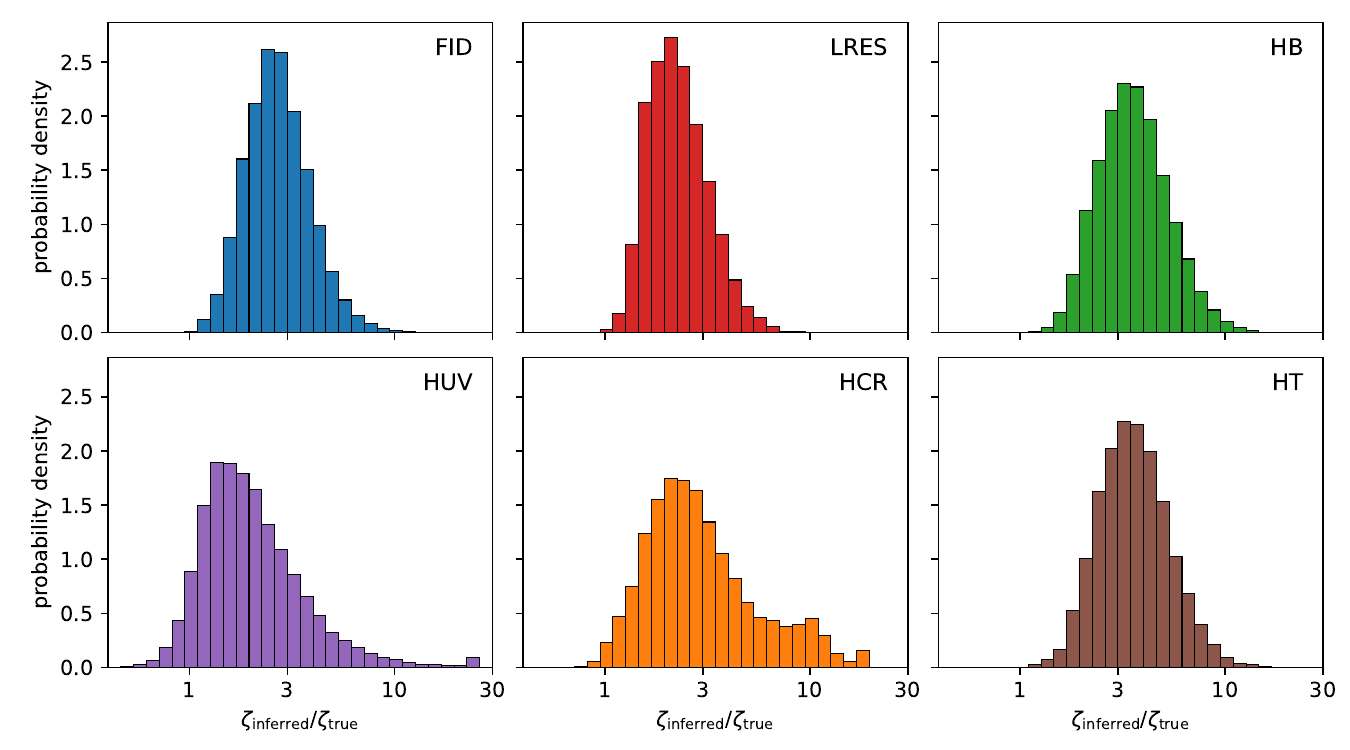}
    \caption{
    Histograms of the inferred-to-true CRIR ratio $\zeta_{\rm inferred}/\zeta_{\rm true}$ for the simulation models varying physical and numerical parameters in Table~\ref{table:parameters}, similar to the fiducial model shown in Figure~\ref{fig:zeta_hist}. The over-estimation of CRIR holds true when varying the physical parameters, with some shifts in the distribution of $\zeta_{\rm inferred}/\zeta_{\rm true}$. The low resolution simulation (LRES) confirms that our numerical resolution is sufficient to achieve convergence in the overall $\zeta_{\rm inferred}/\zeta_{\rm true}$ distribution.}
    \label{fig:zeta_hist_suite}
\end{figure*}

We now examine how the bias in CRIR measurements depends on the physical conditions and numerical resolution for the model suite in Table~\ref{table:parameters}.
Figure~\ref{fig:zeta_hist_suite} shows the distributions of the inferred-to-true CRIR, $\zeta_{\rm inferred}/\zeta_{\rm true}$ for the different simulation models, and Table~\ref{tab:percentiles} shows the median, 25$^{\rm th}$ and 75$^{\rm th}$ percentiles of the distribution in $\zeta_{\rm inferred}/\zeta_{\rm true}$.
All models show $\zeta_{\rm inferred}>\zeta_{\rm true}$ for majority of the sight-lines, with $\zeta_{\rm inferred}/\zeta_{\rm true}$ in the range of $\sim 2-5$. Below, we discuss each model in detail.

\begin{table}[tbp]
    \centering
    \caption{Median, 25th and 75th percentiles of the $\zeta_{\rm inferred}/\zeta_{\rm true}$ distribution for all the simulation models.}
    \label{tab:percentiles}
    \begin{tabular}{lccc}
        \toprule
        \textbf{Model ID} & \textbf{25th Percentile} & \textbf{Median} & \textbf{75th Percentile} \\
        \midrule
        FID  & 2.10 & 2.64 & 3.37 \\
        LRES & 1.77 & 2.21 & 2.82 \\
        HB   & 2.70 & 3.50 & 4.59 \\
        HUV  & 1.35 & 1.84 & 2.71 \\
        HCR  & 1.89 & 2.64 & 4.11 \\
        HT   & 2.71 & 3.52 & 4.63 \\
        \bottomrule
    \end{tabular}
\end{table}

\paragraph{Numerical Resolution}
The lower resolution model \texttt{LRES} shows very similar distributions in $\zeta_{\rm inferred}/\zeta_{\rm true}$ as our fiducial model, confirming that our numerical resolution is sufficient to reach convergence. We have also examined the distribution of $\Ht$ and $\HHHp$ abundances in both models, and found very similar results.

\paragraph{FUV radiation field}
The higher FUV radiation model \texttt{HUV} shows the lowest  $\zeta_{\rm inferred}/\zeta_{\rm true}$ in our simulation suite.
The higher FUV radiation field photo-dissociates ${\rm H_2}$ in low-density gas, confining ${\rm H_2}$ and $\HHHp$ to denser regions where the chemical timescales are shorter, and the chemical abundances are closer to steady-state.

\paragraph{Magnetic field strength}
The higher initial magnetic field run \texttt{HB} slightly increases $\zeta_{\rm inferred}/\zeta_{\rm true}$.
The stronger B-field leads to a higher volume-filling fraction of low-density gas, and therefore enhances the fraction of ${\rm H_2}$ at low densities.

\paragraph{CRIR}
Increasing the CRIR has very little effect on the median value of the $\zeta_{\rm inferred}/\zeta_{\rm true}$ distribution. There is an increase in the number of pixels with very high $\zeta_{\rm inferred}/\zeta_{\rm true}$. 


\paragraph{Turbulent Velocity Dispersion}
The stronger turbulence model with higher velocity dispersion \texttt{HT} gives a higher median value and a marginally wider distribution of $\zeta_{\rm inferred}/\zeta_{\rm true}$ compared to the fiducial model. Stronger turbulence leads to a broader density distribution, resulting in more $\Ht$ in lower-density regions. Higher velocity dispersion also decreases the dynamical time for turbulence advection compared to the chemical time. Both of these effects lead to higher values of $\zeta_{\rm inferred}/\zeta_{\rm true}$.

Overall, our parameter study shows that the CRIR derived from previous analysis of $\HHHp$ observations are over-estimated by a factor of $\sim 2-5$ due to the assumption of steady-state chemistry. This bias is more severe in sight-lines with lower gas column density, lower FUV radiation field or CRIR, stronger magnetic field strengths, and stronger turbulence.






\section{Discussion}\label{sec:Discussion}
In this section, we describe the impact of our results on our understanding of the mean and dispersion of the CRIR in nearby molecular clouds.
\subsection{Revised estimate of the mean CRIR}
\citet{Obolentseva2024} and \citet{Indriolo2026} inferred the CRIR in 16 sight-lines under the assumption of steady-state chemistry.  To quantify the importance of time-dependent chemistry on their results, we applied the following procedure. For each of their 16 sight-lines, we tabulated the column density $N_i$, and the reported ionization rate $\zeta_i$.  For each $N_i$, we used the results of the present paper to calculate the mean value $R_i$ of $\zeta_{\rm inferred}/\zeta_{\rm true}$, and the mean value $\tilde R_i$ of $\zeta_{\rm true}/\zeta_{\rm inferred}$.  $R_i$ and $\tilde R_i$ are the average values of the corresponding quantities from the $3 \times 256^2$ sight-lines each with column density $N_j^{\rm sim}$ in our fiducial simulation, weighted by the function
\begin{equation}
    W(N_j^{\rm sim}, N_i) = \exp\left[-(\log_{10} \left[N_j^{\rm sim}\right] - \log_{10} \left[N_i\right])^2 / (2\sigma^2)\right],
    \label{eq:weight}
\end{equation}
where $\sigma = 0.1$   We note that our choice of $\sigma = 0.1$ is comparable to the typical observational uncertainty on $\log_{10}{(N_i}$ \citep{Indriolo2026}.  We also tried  $\sigma = 0.03$ and found changes of less than 2\% to the values of $\bar \zeta_{\rm true}$ and $\zeta_{\rm const}$ reported in Equations \eqref{eq:zeta_true_corrected} and \eqref{eq:zeta_const_corrected} as well as the median value of the standard deviation of our model inferred ionization rates (see paragraph following Equation \eqref{eq:zeta_const_corrected}). 
\par
We estimate the true mean ionization rate per ${\rm H_2}$ as, 
\begin{equation}\label{eq:zeta_true_corrected}
    \bar \zeta_{\rm true} = \frac{1}{16} \sum_{i = 1}^{16} \zeta_i \tilde R_i = 2.1 \times 10^{-17} \, {\rm s}^{-1}.
\end{equation}

\munan{We reiterate that the estimate in Equation \eqref{eq:zeta_true_corrected} assumes that our fiducial simulation represents the typical conditions in the local ISM. As discussed in \ref{sec:param_dependence}, $\zeta_{\rm inferred}/\zeta_{\rm true}$ depends on the environmental conditions.}

\subsection{Dispersion in inferred CRIR attributable to non-equilibrium chemistry}
We also considered a model in which the true ionization rate per ${\rm H_2}$ for all sight-lines is equal to a constant value, $\zeta_{\rm const}$.  We chose $\zeta_{\rm const}$ so that the expectation of the inferred ionization rate would be equal to $\bar \zeta_i = 5.3 \times 10^{-17}$ s$^{-1}$, as reported in \citet{Indriolo2026}:
\begin{equation}\label{eq:zeta_const_corrected}
    \zeta_{\rm const} = \frac{\bar \zeta_i}{ \frac{1}{16} \sum_{i = 1}^{16} R_i} = 1.94 \times 10^{-17} {\rm s}^{-1}.
\end{equation}
We then produced $10^4$ sets of 16 model-inferred ionization rates.  To construct each of the $10^4$ sets, we multiplied $\zeta_{\rm const}$ by a randomly selected sample of 16 values of $\zeta_{\rm inferred}/\zeta_{\rm true}$ from our fiducial simulation.  The probability of selecting a given value is proportional to the weight given in Equation \eqref{eq:weight}, and therefore depends on the observed column density for that sight-line.  We calculate the standard deviation for each set of model-inferred ionization rates.  We find a median value of $1.9 \times 10^{-17}$ s$^{-1}$, as compared with the observed standard deviation of $2.5 \times 10^{-17}$ s$^{-1}$ in the sample from \citet{Obolentseva2024} and \citet{Indriolo2026}.  This suggests that more than half of the dispersion may arise from variations in the value of $\zeta_{\rm inferred}/\zeta_{\rm true}$ caused by time-dependent chemistry.

\subsection{Comparison with the Voyager Measurements}
Our estimated true mean ionization rate is slightly below what would be produced by the local cosmic ray energy spectrum measured by the Voyager probes \citep{Cummings2016, Stone2019}.  \citet{Cummings2016} estimated a primary ionization rate of atomic hydrogen of $\zeta_p=(1.01 - 1.09) \times 10^{-17}$ s$^{-1}$.  This estimate is a lower bound, since it neglects the contribution of particles with energies less than 3 MeV (the lowest energy measured with Voyager). Using the standard relation \citep{Glassgold1974} that the total ionization rate of molecular hydrogen is 2.3 times the primary ionization rate of atomic hydrogen, this corresponds to an ionization rate of approximately $\zeta_{\rm H_2} = 2.4 \times 10^{-17}$ s$^{-1}$.  If we use the more accurately determined ratio of secondary to primary ionization of 1.0 \citep{Ivlev2021}, and keep the ratio of the ionization cross-section of molecular hydrogen to that of atomic hydrogen equal to 1.53 \citep{Glassgold1974}, then we obtain a value of approximately $\zeta_{\rm H_2} = 3.1 \times 10^{-17}$ s$^{-1}$.  We reiterate that this is a lower bound.  Using a spectrum similar to Voyager, and taking into account shielding by intervening material \kedron{using the ballistic propagation model in \citet{Padovani2018},} \citep{Indriolo2026} find an ionization rate between 3 and 5 $\times 10^{-17}$ s$^{-1}$ at column densities between 1 and 6 $\times 10^{21}$ cm$^{-2}$.
\subsection{Implications for cosmic ray attenuation in molecular clouds}
Because the correction factor $\zeta_{\rm inferred}/\zeta_{\rm true}$ is typically larger for smaller values of the column density (see Figure \ref{fig:violin}), the hint of a trend towards reduced ionization rate at higher column density in \citet{Indriolo2026} goes away if we account for this.  We use least-squares regression to fit a straight line to a plot of $\log\left(\zeta_i \tilde R_i \right)$ vs $\log N_i$, finding a best-fitting slope of 0.1.  The corresponding slope in the best-fit line to $\log\left(\zeta_i\right)$ vs $\log N_i$ is -0.5.  We did not include error bars on either $\zeta_i$ or $N_i$ in this fit.  Given the many additional difficult-to-quantify uncertainties, we do not aim to make a quantitative statement regarding the variation of the CRIR with column density.  We do point out that the inclusion of non-equilibrium chemistry may weaken evidence for shielding of CRs in molecular clouds from analysis of ${\rm H_3^+}$ observations.  \kedron{If we are right that the mean CRIR is below what the Voyager spectrum predicts when combined with the ballistic propagation model, this suggests either that the CR spectrum measured by Voyager is higher than that in the vicinity of nearby molecular clouds, or that some transport phenomenon is efficiently preventing CRs from entering the molecular clouds.  Scattering off of pre-existing \citep{Silsbee2019} or self-generated \citep{Chernyshov2025} turbulence has been argued to reduce the abundance of low-energy CRs in molecular clouds, however this would also result in a steeper dependence of the CRIR on column density, which is not supported by our results.  Given the remaining uncertainties and the small range of column density probed by these measurements, we do not feel confident in making any strong statements regarding the implications of these results for CR transport physics. }
\section{Conclusions}  \label{sec:conclusions}
In this paper, we investigate the effect of time-dependent chemistry on the CRIR inferred from $\HHHp$ observations. We perform 3D MHD simulations of the atomic and molecular ISM with coupled chemistry and driven turbulence. To assess the assumption of steady-state chemistry adopted in the original analysis of  $\HHHp$ observations \citep{Obolentseva2024, Indriolo2026}, we post-process our simulations with time-dependent chemistry and construct steady-state chemical abundances over a range of CRIR values. Mock CRIR measurements are obtained along each simulation sightline based on the observational procedure of selecting the CRIR for which the steady-state abundance ratio $N_\HHHp/N_\Ht$ matches the time-dependent (``observed'') value. This inferred CRIR ($\zeta_\mathrm{inferred}$) is then compared to the ``true'' CRIR ($\zeta_\mathrm{true}$) adopted in the time-dependent simulations to quantify the observational bias. Our main findings are summarized as follows:
\begin{enumerate}
    \item The abundances of both $\HHHp$ and $\Ht$ are higher in time-dependent chemistry simulations than the steady-state values (Figures \ref{fig:map} and \ref{fig:2dhist}), with the difference being most pronounced in low-density regions ($n\lesssim100~\mathrm{cm^{-3}}$). This enhancement is caused by the advection of $\Ht$ from dense to diffuse gas, combined with the efficient self-shielding of $\Ht$. Due to its short reaction timescales, $\HHHp$ remains in almost perfect chemical equilibrium with $\Ht$.
   
    \item The column density ratio $N_\HHHp/N_\Ht$ is higher in time-dependent chemistry simulations than in the steady-state case (Figure \ref{fig:NH3_over_NH2}).
    \kawai{This is a direct consequence of the enhanced $\Ht$ abundance in diffuse gas,}
    \munan{where the lower recombination rate for $\HHHp$ leads to a higher abundance ratio of $f_\HHHp/f_\Ht$.}
    
    \item The inferred CRIR under the steady-state assumption is a factor of $\sim 2-5$ higher than the true CRIR in time-dependent chemistry simulations (Figures \ref{fig:zeta_hist} and \ref{fig:zeta_hist_suite}), with a median value of $\zeta_\mathrm{inferred}/\zeta_\mathrm{true} \approx 3$ (Table \ref{tab:percentiles}). This implies that properly accounting for time-dependent chemistry yields both a lower average value and a smaller dispersion of CRIR from observations.
    
    \item $\zeta_\mathrm{inferred}/\zeta_\mathrm{true}$ is higher along sightlines with lower column densities, where the time-dependent effect is more prominent (Figure \ref{fig:violin}).
    \item $\zeta_\mathrm{inferred}/\zeta_\mathrm{true}$ increases with stronger magnetic fields, weaker FUV radiation field, and stronger turbulence (Figure \ref{fig:zeta_hist_suite}). Our fiducial simulation produces a distribution of $\zeta_\mathrm{inferred}/\zeta_\mathrm{true}$ similar to that in the lower-resolution run, demonstrating numerical convergence.

    \item 
    \munan{Using our fiducial model,}
    we \kawai{estimate} an average CRIR per $\Ht$ of $\zeta_\Ht \approx 2\times 10^{-17}~\mathrm{s^{-1}}$ from the $\HHHp$ observations, lower than the value given by the original analyses in \citet{Obolentseva2024} and \citet{Indriolo2026}. 
    \munan{We note that the correction $\zeta_\mathrm{inferred}/\zeta_\mathrm{true}$ and the implied CRIR depend on the assumed environmental conditions of the ISM (Section \ref{sec:param_dependence}).}
    Our estimate is \munan{slightly below} that implied by the Voyager measurement of the CR spectrum \citep{Cummings2016, Padovani2018}.
    
    \item Time-dependent chemistry can account for most of the observed dispersion in the CRIR, as well as the trend with column density in the original analysis under the steady-state assumption. The resulting CRIR is consistent with a constant value in the column density range of $N=(2-6)\times10^{21}~\mathrm{cm^{-2}}$.
\end{enumerate}


\section*{Acknowledgments}
This research used resources of the National Energy Research Scientific Computing Center, a DOE Office of Science User Facility supported by the Office of Science of the U.S. Department of Energy under Contract No. DE-AC02-05CH11231, using NERSC awards ASCR-ERCAP0036667 and FES-ERCAP0036102.
This research was supported 
by grants 216179 from the Simons Foundation and NSF PHY-2309135 to the Kavli Institute
for Theoretical Physics (KITP).
\munan{We thank the anonymous referee for a constructive review, which helped to improve the overall quality and clarity of the paper.}



\bibliographystyle{aasjournal}
\bibliography{all}



\end{document}